\documentstyle[fleqn,twoside,epsf]{article} 
\newcommand{\be}{ \begin{eqnarray}}
\newcommand{\ee}{\end{eqnarray}}
\newcommand{\beno}{ \begin{eqnarray*}}
\newcommand{\eeno}{\end{eqnarray*}}
\newcommand{\raf}[1]{(\ref{#1})}

\begin{document}

\pagestyle{empty}
\markboth{\rm KO, KOCH \& LI}{\rm IN-MEDIUM EFFECTS}

\hfill {\large LBNL-39866} \\
\ \\

\noindent
{\huge PROPERTIES OF HADRONS IN THE NUCLEAR MEDIUM}
\bigskip

\noindent
{\large {\it Che Ming Ko\footnote{e-mail:ko@comp.tamu.edu}}}

\medskip
\noindent
Cyclotron Institute and Physics Department\\
Texas A\&M University, College Station, Texas 77843, USA

\medskip
\noindent
{\large {\it Volker Koch\footnote{e-mail:vkoch@lbl.gov}}}

\medskip
\noindent 
Nuclear Science Division\\
Lawrence Berkeley National Laboratory,
Berkeley, CA 94720, USA

\medskip
\noindent
{\large {\it Guoqiang Li\footnote{e-mail:gqli@nuclear.physics.sunysb.edu}}}

\medskip
\noindent 
Department of Physics\\
State University of New York at Stony Brook, 
Stony Brook, N.Y. 11794, USA

\bigskip
\noindent
{\sc key words}:\quad  Chiral symmetry, Goldstone bosons, vector mesons,
baryons, hot and dense hadronic matter, heavy-ion collisions

\bigskip
\hrule
\bigskip
\begin{abstract}

This review is devoted to the discussion of hadron properties in the nuclear
medium and its relation to the partial restoration of chiral symmetry. 
Special attention is given to disentangle in-medium effects due to 
conventional many-body interactions from those due to the change of the chiral 
condensate. In particular, we shall discuss medium effects on the Goldstone 
bosons (pion, kaon and eta), the vector mesons (rho, omega, phi), and 
the nucleon. Also, for each proposed in-medium effect the experimental 
consequence and results will be reviewed.

\end{abstract}

\bigskip
\hrule
\bigskip
\newpage
\pagestyle{myheadings}
\tableofcontents

\bigskip

\section{INTRODUCTION} 

The atomic nucleus provides a unique laboratory to study the long range and
bulk properties of QCD. Whereas QCD is well tested in the perturbative regime
rather little is known about its 
properties in the long range, nonperturbative region. One of the central 
nonperturbative properties of QCD is the spontaneous breaking chiral 
symmetry in the ground state resulting in a nonvanishing scalar quark 
condensate, $<\bar{q} q > \neq 0$.  It is believed and supported by lattice
QCD calculations \cite{DeT95} that at temperatures around $150 \, \rm MeV$,
QCD undergoes a phase transition to a chirally restored phase, characterized 
by the vanishing of the order parameter, the chiral condensate 
$<\bar{q} q >$. This is supported by results 
obtained within chiral perturbation theory \cite{DL90,CFG92}. 
Effective chiral models predict that a similar transition  
also takes place at finite nuclear density. 

The only way to create macroscopic, strongly interacting systems at finite 
temperature and/or density in the laboratory is by colliding heavy nuclei 
at high energies. Experiments carried out at various bombarding energies, 
ranging from 1 AGeV (BEVALAC, SIS) to 200 AGeV (SPS), have established that 
one can generate systems of large density but moderate temperatures 
(SIS,BEVALAC), systems of both large density and temperature (AGS) as 
well as systems of low density and high temperatures (SPS). Therefore, a 
large region of the QCD phase diagram can be investigated through the 
variation of the bombarding energy. But in addition the atomic nucleus 
itself represents a system at zero temperature and finite density. At 
nuclear density the quark condensate is estimated to be reduced by 
about 30\% \cite{DL90,CFG92,LK94,BW96} 
so that effects due to the change of the 
chiral order parameter may be measurable in reactions induced by a pion, 
proton or photon on the nucleus. 

Calculations within the instanton liquid model \cite{SS96} as well as
results from phenomenological models for hadrons \cite{Rho94} suggest
that the properties of the light hadrons, such as masses and couplings, are
controlled by chiral symmetry and its spontaneous as well as explicit breaking. 
Confinement seems to play a lesser role. If this is the correct picture of 
the low energy excitation of QCD, hadronic properties should depend on the 
value of the chiral condensate $<\bar{q} q >$. Consequently, 
we should expect that the properties of hadrons change considerably 
in the nuclear environment, where the chiral condensate is reduced. 
Indeed, based on the restoration of scale invariance of QCD, Brown and 
Rho have argued that masses of nonstrange hadrons would scale with the
quark condensate and thus decrease in the nuclear medium \cite{BR91}. 
This has since stimulated extensive theoretical and experimental
studies on hadron in-medium properties.

By studying medium effects on hadronic properties one can directly test our 
understanding of those non-perturbative aspects of QCD, 
which are responsible for the light hadronic states.  The best way to 
investigate the change of hadronic properties in experiment is to study 
the production of particles, preferably photons and dileptons, as they
are not affected by final-state interactions. Furthermore, since vector mesons
decay directly into dileptons, a change of their mass can be seen directly in
the dilepton invariant mass spectrum. In addition, as we shall discuss, 
the measurement of subthreshold particle production such as kaons 
\cite{AK85} and antiprotons \cite{BCM91} may also reveal some rather 
interesting in-medium effects.

Of course a nucleus or a hadronic system created in relativistic heavy ion
collisions are strongly interacting. Therefore, many-body excitations can carry
the same quantum numbers as the hadrons under consideration and thus can mix
with the hadronic states. In addition, in the nuclear environment `simple' 
many-body effects such as the Pauli principle are at work, which, as we 
shall discuss, lead to considerable modifications of hadronic properties
in some cases. How these effects are related to the partial restoration 
of chiral symmetry is a new and unsolved question in nuclear many-body
physics.  One example is the effective mass of a nucleon in the medium,
which was introduced long time ago \cite{JLM76,MBB85} to model the momentum 
dependence of the nuclear force, which is due to its finite range, or to 
model its energy dependence, which results from higher order (2-particle 
- 1-hole etc.) corrections to the nucleon self energy. On the other hand,
in the relativistic mean-field description one also arrives at a reduced 
effective mass of the nucleon. According to \cite{BWB87}, this is due to 
so-called virtual pair corrections and may be related to chiral symmetry 
restoration as suggested by recent studies \cite{CFG92,GR95,BR96,FTS95}. 

Another environment, which is somewhat `cleaner' from the theorists 
point of view, is a system at finite temperature and vanishing 
baryon chemical potential. At low temperatures such a system can be 
systematically explored within the framework of chiral perturbation 
theory, and essentially model-independent statements about the
effects of chiral restoration on hadronic masses and couplings may thus 
be given. In the high temperature regime eventually 
lattice QCD calculations should be able to tell us about the properties 
of hadrons close to the chiral transition temperature. Unfortunately, 
such a system is very difficult to create in the laboratory.

Since relativistic heavy ion collisions are very complex processes, one has 
to resort to careful modeling in order to extract the desired in-medium 
correction from the available experimental data.
Computationally the description of these collisions is best
carried out in the framework of transport theory 
\cite{SG86,BD88,CMM90,AIC91,BKM93,BGM94,KL96} since nonequilibrium effects
have been found to be important at least at energies up to 2 AGeV. At higher
energies there seems to be a chance that one can separate the reaction in an
initial hard scattering phase and a final equilibrated phase. It appears that
most observables are dominated by the second stage of the reaction so that one
can work with the assumption of local thermal equilibrium. However, as compared
to hydrodynamical calculations, also here the transport approach has some
advantages since the freeze out conditions are determined from the calculation
and are not needed as input parameter. As far as a photon or a proton induced
reaction on nuclei is concerned, one may resort to standard reaction theory. 
But recent calculations seem to indicate that these reaction can also
be successfully treated in the transport approach 
\cite{EHT96a,EHT96b,LBK96}. Thus it appears possible that one can explore 
the entire range of experiments within one and the same theoretical framework 
which of course has the advantage that one reduces the ambiguities of the 
model to a large extent by exploring different observables.

This review is devoted to the discussion of in-medium effects in the hadronic
phase and its relation to the partial restoration of chiral symmetry. We,
therefore, will not discuss another possible in-medium effect which is related
to the deconfinement in the Quark Gluon Plasma, namely the suppression of the
$J/\Psi$. This idea, which has been first proposed by Matsui and Satz
\cite{MS86} is based on the observation that due to the screening of
the color interaction in the Quark Gluon Plasma, the $J/\Psi$ is not bound
anymore. As a result, if such a Quark Gluon Plasma is formed in a
relativistic heavy ion collision, the abundance of $J/\Psi$ should be
considerably reduced. Of course also this signal suffers from more conventional
backgrounds, namely the dissociation of the $J/\Psi$ due to hadronic
collisions \cite{GH92,Gav90,CK96}. To which extent present data can be 
understood in a purely hadronic scenario is extensively debated at the 
moment \cite{QM96}. We refer the reader to the literature for further 
details \cite{KS95}.

In this review we will concentrate on proposed in-medium effects on
hadrons. Whenever possible, we will try to disentangle conventional in-medium 
effects from those we believe are due to new physics, namely the change of 
the chiral condensate. We first will discuss medium effects on the Goldstone
bosons, such as the pion, kaon and eta. Then we will concentrate on the vector
mesons, which have the advantage that possible changes in their mass can be
directly observed in the dilepton spectrum. We further discuss the
effective mass of a nucleon in the nuclear medium.
Finally, we will close by summarizing the current status of Lattice QCD
calculations concerning hadronic properties. As will become clear from our 
discussion, progress in this field requires the input from experiment. 
Many question cannot be settled from theoretical consideration alone. 
Therefore, we will always try to emphasize the observational aspects for 
each proposed in-medium correction.

\section{GOLDSTONE BOSONS} 

This first part is devoted to in-medium effects of Goldstone bosons,
specifically the pion, kaon and eta. The fact that these particles 
are Goldstone bosons means that their properties are directly linked to the
spontaneous breakdown of chiral symmetry. Therefore, rather reliable
predictions about their properties can be made using chiral symmetry arguments
and techniques, such as chiral perturbation theory. Contrary to naive
expectations, the properties of Goldstone bosons are rather robust with 
respect to changes of the chiral condensate. On second thought, however, 
this is not so surprising, because as long as chiral symmetry is spontaneously 
broken there will be Goldstone bosons. The actual nonvanishing value of 
their mass is due to the explicit symmetry breaking, which -- at least in 
case of the pion -- are small.  Changes in their mass, therefore, are 
associated with the sub-leading explicit symmetry breaking terms and
are thus small.  We should stress, however, that in case of the kaon
the symmetry breaking terms are considerably larger, leading to sizeable 
corrections to their mass at finite density as we shall discuss below. 

\subsection{\it The pion}
\label{pion}
Of all hadrons, the pion is probably the one where in-medium corrections 
are best known and understood.  
From the theoretical point of view, pion properties can be well determined
because the pion is such a `good' Goldstone boson. The  explicit 
symmetry breaking terms in this case are small, as they are associated with the
current masses of the light quarks. Therefore, the properties of the 
pion at finite density and temperature can be calculated using for instance 
chiral perturbation theory. But more importantly, there exists a considerable 
amount of experimental data ranging from pionic atoms to pion-nucleus
experiments to charge-exchange reactions. These data all address the pion
properties at finite density which we will discuss now.

\subsubsection*{\sc Finite density}
The mass of a pion in symmetric nuclear matter is directly related to the
real part of the isoscalar-s-wave pion optical potential. This has been 
measured to a very high precision in pionic atoms \cite{EW88}, and one
finds to first order in the nuclear density $\rho$
\be
\Delta m_\pi^2 = - 4 \pi (b_0)_{\rm eff} \rho, \,\,\, (b_0)_{\rm eff} 
\simeq -0.024 m_\pi^{-1}.
\ee
At nuclear matter density the shift of the pion mass is 
\be
\frac{\Delta m_\pi}{m_\pi} \sim 8 \%,
\ee
which is small as  expected from low energy theorems based on chiral
symmetry. Theoretically, the value of $(b_0)_{\rm eff}$ is well understood 
as the combined contribution from a single scattering (impulse approximation) 
involving the small s-wave isoscalar $\pi N$ scattering length 
$b_0 = -0.010(3) m_\pi^{-1}$ and the contribution from a (density dependent) 
correlation 
or re-scattering term, which is dominated by the comparatively large 
isovector-s-wave scattering length $b_1 = -0.091(2)m_\pi^{-1}$, i.e., 
\be
(b_0)_{\rm eff} = b_0 - [1 + \frac{m_\pi}{M_N} (b_0^2 + 2 b_1^2)] 
<\frac{1}{r}>,
\ee
where the correlation length $<1/r> \simeq 3p_f/(2\pi)$ is essentially 
due to the Pauli exclusion principle.  The weak and slightly repulsive 
s-wave pion potential has also been found in calculations based on 
chiral perturbation theory \cite{DEE92,TW95} as well as the phenomenological 
Nambu - Jona-Lasinio model \cite{NJL}. This is actually not too surprising 
because both amplitudes, $b_0$ and $b_1$, are controlled by 
chiral symmetry and its explicit breaking \cite{EW88,Wei66}.

Pions at finite momenta, on the other hand, interact very strongly with nuclear
matter through the p-wave interaction, which is dominated by the $P_{33}$
delta-resonance. This leads to a strong mixing of the pion with nuclear
excitations such as a particle-hole and a delta-hole excitation. As a result 
the pion-like excitation spectrum develops several branches, which to 
leading order correspond to the pion, particle-hole, and delta-hole
excitations. Because of the attractive p-wave interaction the
dispersion relation of the pionic branch becomes considerably softer 
than that of a free pion. A simple model, which has been used in practical 
calculations \cite{BBK88,XKK93} is the so-called delta-hole model 
\cite{EW88,FPU81} which concentrates on the stronger pion-nucleon-delta 
interaction ignoring the nucleon-hole excitations. In this model, the 
pion dispersion relation in nuclear medium can be written as 
\begin{eqnarray}
\omega ({\bf k}, \rho) =m_\pi^2+{\bf k}^2+\Pi (\omega, {\bf k}),
\end{eqnarray}
where the pion self-energy is given by
\begin{eqnarray}
\Pi (\omega, {\bf k})={{\bf k}^2 \chi(\omega,{\bf k})\over 1-g^\prime\chi
(\omega,{\bf k})},
\end{eqnarray}
with $g^\prime\approx 0.6$ the Migdal parameter which accounts for short-range
correlations, also known as the Ericson-Ericson --
Lorentz-Lorenz effect \cite{EE66}. The pion susceptibility $\chi$ is given by
\begin{eqnarray}
\chi(\omega , {\bf k}) \approx {8\over 9}\Big({f_{\pi N\Delta} \over m_\pi
}\Big)^2 {\omega_R \over \omega ^2-\omega^2_R}
{\rm exp}\Big(-2{\bf k}^2/b^2\Big) \rho,
\end{eqnarray}
where $f_{\pi N\Delta}\approx 2$ is the pion-nucleon-delta coupling 
constant,
$b\approx 7m_\pi$ is the range of the form factor, and
$\omega_R\approx {{\bf k}^2\over 2m_\Delta } +m_\Delta -m_N$.
 
\begin{figure}[htp]
\setlength{\epsfxsize=0.7}{\textwidth}
\centerline{\epsffile{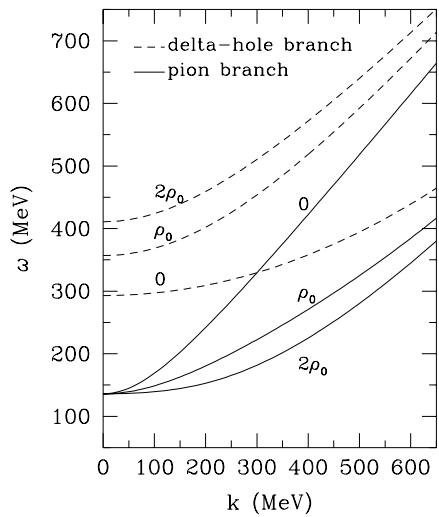}}
\caption{Pion dispersion relation in the nuclear medium. The normal
nuclear matter density is denoted by $\rho_0$.}
\label{fig_pion_disp}
\end{figure}

The pion dispersion relation obtained in this model is shown in Fig. 
\ref{fig_pion_disp}. The pion branch in the lower part of the
figure is seen to become softened, while the delta-hole branch in the
upper part of the figure is stiffened.  Naturally this model oversimplifies 
things and once couplings to nucleon-hole excitations and corrections to 
the width of the delta are consistently taken into account, the strength 
of the delta-hole branch is significantly reduced 
\cite{BOV89,KXS89,XSS94,HU94,KM95,HR95}. However, up to momenta of about 
$2 m_\pi$ the pionic branch remains pretty narrow and considerably softer 
than that of a free pion. 
The dispersion relation of the pion can be measured directly using 
$(\mbox{}^3He,t)$ charge exchange reactions \cite{HRB93}. In these experiments
the production of coherent pion have been inferred from angular correlations
between the outgoing pion and the transferred momentum. The corresponding 
energy transfer is smaller than that of free pions indicating
the attraction in the pion branch discussed above. More detailed measurements
of this type are currently being analyzed \cite{gaarde}.

In the context of relativistic heavy ion collisions interest in the in-medium 
modified pion dispersion relation got sparked by the work of Gale and 
Kapusta \cite{GK87}. Since the density of states is proportional to the 
inverse of the group velocity of these collective pion modes, which 
can actually vanish depending on the strength of the interaction (see
Fig. \ref{fig_pion_disp}), they  argued that a softened dispersion 
relation would lead to a strong enhancement of the dilepton yield.
This happens at invariant mass close to twice the pion mass because the 
number of pions with low energy will be considerably increased. 
However, as pointed out by Korpa and Pratt \cite{KP90} and further explored in
more detail in \cite{KXK90}, the initially proposed strong enhancement is  
reduced considerably once corrections due to gauge invariance are properly 
taken into account. 
Other places, where the effect of the pion dispersion relation is expected to
play a role, are the inelastic nucleon-nucleon cross section \cite{BBK88} 
and the shape of the pion spectrum \cite{XKK93,ECE93,FSL96}. In the latter 
case one finds \cite{XKK93} that as a result of the attractive interaction 
the yield of pions at low $p_t$ is enhanced by about a factor of 2. This low 
$p_t$ enhancement is indeed seen in experimental data for $\pi^-$ production
from the BEVALAC \cite{Ody87} and in more recent data from the TAPS 
collaboration \cite{Ber90}, which measures neutral pions.  Certainly, in 
order to be conclusive, more refined calculations are needed, and they are 
presently being carried out \cite{HR95}. 

\subsubsection*{\sc Finite temperature}
The pionic properties at finite temperature instead of finite density are  
qualitatively very similar, although the pion now interacts mostly with 
other pions instead of nucleons.  Again, due to chiral symmetry, the s-wave 
interaction among pions is small and slightly repulsive, leading to a small 
mass shift of the pion \cite{GL89}. 

And analogous to the case of nuclear matter, 
there is a strong attractive p-wave 
interaction, which is now dominated by the $\rho$-resonance. 
This again results 
in a softened dispersion relation which, however, is not as dramatic as 
that obtained at finite density \cite{Shu90,Shu91,KB93,Son94}.  Also the 
phenomenological consequences are similar. The modified dispersion relation 
results in an enhancement of dileptons from the pion annihilation by a factor 
of about two at invariant masses around $300 - 400 \, \rm MeV$
\cite{SKL96,KS96,SK96}. Unfortunately, in the same mass range other channels
dominate the dilepton spectrum (see discussion in section \ref{rho}) so that
this enhancement cannot be easily observed in experiment \cite{KS96}.  
The modified dispersion relation has also been invoked in order to explain the 
enhancement of low transverse momentum pions observed at CERN-SPS heavy ion 
collisions \cite{Shu91,Str89,Str90}. However, at these high energies the 
expansion velocity of the system, which is mostly made out of pions, is too 
fast for the attractive pion interaction to affect the pion spectrum 
at low transverse momenta \cite{KB93}. This is different at the lower energies
around $1 \, \rm GeV$. There, the dominant part of the system and the source 
for the pion potential are the nucleons, which move considerably slower than
pions. Therefore, pions have a chance to leave the potential well before 
it has disappeared as a result of the expansion.  

To summarize this section on pions, the effects for the pions at finite 
density as well as finite temperature are dominated by p-wave resonances, the
$\Delta(1230)$ and the $\rho(770)$, respectively. Both are not related directly
to chiral symmetry and its restoration but rather to what we call here 
many-body effects, which of course does not make them less interesting. 
The s-wave interaction is small because the pion is such a `good' Goldstone 
Boson; probably too small to have any phenomenological consequence (at 
least for heavy ion collisions). 

\subsection{\it Kaons}

Contrary to the pion, the kaon is not such a good Goldstone boson. Effects of
the explicit chiral symmetry breaking are considerably bigger, as one can see
from the mass of the kaon, which is already half of the typical hadronic mass 
scale of $1 \, \rm GeV$. In addition, since the kaon carries strangeness, its
behavior in non-strange, isospin symmetric matter will be different from that
of the pion. Rather interesting phenomenological consequences arise from this
difference such as a possible condensation of antikaons in neutron star matter
\cite{KN86,NK87,PW91,BLR94}.

\subsubsection*{\sc Chiral Lagrangian}

This difference can best be exemplified by studying the leading order 
effective $\rm SU(3)_L \times SU(3)_R$ Lagrangian obtained in heavy baryon
chiral perturbation theory \cite{BLR94,JM92}
\begin{eqnarray}
{\cal L}_0&=&{f^2\over 4} \, {\rm Tr}~\partial^\mu U\partial_\mu U^+
+\frac{f^2}{2}\, r \,{\rm Tr}~M_q(U + U^+ - 2)
\nonumber \\
&& + {\rm Tr}~\bar{B} i \, v_\mu {\cal D}^\mu B
+2 D \, {\rm Tr}~\bar{B} S^\mu \{A_\mu,B\}
\nonumber \\
&& + 2 F \,  {\rm Tr}~\bar{B}S^\mu [A_\mu,B].
\label{heavy_baryon_0}
\end{eqnarray}
In the above formula, we have $U=\exp(2i\pi/f)$ with $\pi$ and $f$ being the 
pseudoscalar meson octet and their decay constant, respectively; $B$ is 
the baryon octet; $v_\mu$ is the four velocity of the heavy baryon 
($v^2 = 1$); and $S_\mu$ stands for the spin operator $S_\mu = 
\frac14 \gamma_5 [v_\nu \gamma_\nu, \gamma_\mu]$ and $v_\mu S^\mu = 0$;
$M_q$ is the quark mass matrix; and $r$, $D$ as well as $F$ are 
empirically determined constants. Furthermore,
\be
{\cal D_\mu B} &=& \partial_\mu B + [V_\mu,B]
\\
V_\mu &=& \frac12 (\xi \partial_\mu \xi^\dagger + 
\xi^\dagger \partial_\mu \xi),
\,\,\,\, 
A_\mu = \frac{i}{2} 
(\xi \partial_\mu \xi^\dagger - \xi^\dagger \partial_\mu \xi),
\ee
with $\xi^2 = U$. 

To leading order in the chiral counting the explicit symmetry breaking term 
$\sim {\rm Tr}~ M_q(U + U^+ - 2)$ gives rise to the masses of the Goldstone
bosons. The interesting difference between the behavior of pions and kaons in
matter arises from the term involving the vector current $V_\mu$, i.e.
${\rm Tr}~ \bar{B} i v_\mu {\cal D}^\mu B$. In case of the pion, this term 
is identical to the well-known Weinberg-Tomozawa term \cite{Wei66}
\be
\delta {\cal L}_{WT} = \frac{-1}{4 f_\pi^2} (\bar{N} \vec{\tau} \gamma^\mu N)
\cdot\left(\vec{\pi} \times (\partial_\mu \vec{\pi})\right).
\ee
It contributes only to the isovector s-wave scattering amplitude and, 
therefore, does not contribute to the pion optical potential in isospin 
symmetric nuclear matter. This is {\em different} in case of the kaon,
where we have
\be
\delta {\cal L}_{WT} = \frac{-i}{8 f^2} \left( 3 (\bar{N} \gamma^\mu N) 
(K\buildrel\leftrightarrow\over 
\partial_\mu K)+ (\bar{N} \vec{\tau} \gamma^\mu N) 
(K\vec{\tau} \buildrel\leftrightarrow\over \partial_\mu K) \right).
\ee 
The first term contributes to the isoscalar s-wave amplitude and, 
therefore, gives rise to an attractive or repulsive optical potential 
for $K^-$ and $K^+$ in symmetric nuclear matter. It turns out, however, that
this leading order Lagrangian leads to an s-wave scattering length which is too
repulsive as compared with experiment. Therefore, terms next to leading order
in the chiral expansion are needed. Some of these involve the
kaon-nucleon sigma term and thus are sensitive to the second difference between
pions and kaons, namely the strength of the explicit symmetry breaking. 
The next to leading order effective kaon-nucleon Lagrangian can be written as
\cite{BLR94} 
\be
{\cal L}_{\nu=2} = \frac{\Sigma_{KN}}{f^2} \left( \bar{N} N \right) 
(\bar{K}K) 
+\frac{C}{f^2} \left( \bar{N} \vec{\tau} N \right) \cdot 
\left(\bar{K}\vec{\tau} K \right)  
\nonumber \\
+\frac{\tilde{D}}{f^2} \left( \bar{N} N \right) 
(\partial_t \bar{K} \partial_t K)                
+\frac{\tilde{D}'}{f^2} \left( \bar{N} \vec{\tau}  N \right) \cdot 
\left( \partial_t \bar{K}\vec{\tau} \partial_t K \right).
\ee
The value of the kaon-nucleon sigma-term $\Sigma_{KN} = \frac12 (m_q + m_s)
\langle N |  \bar{u} u + \bar{s}s | N \rangle$ depends on the strangeness
content of the nucleon, $y=2\langle N|\bar ss|N\rangle \\
/\langle N|\bar uu+\bar dd|N\rangle$ $\approx 0.1-0.2$. Using the light quark
mass ratio $m_s/m\approx 29$, one obtains $370<\Sigma_{KN}<405 \, \rm
MeV$. The additional parameters, $C, \tilde{D}, \tilde{D}'$ are then fixed by
comparing with $K^+$-nucleon scattering data \cite{BLR94}. This so determined
effective Lagrangian can then be used to predict the $K^-$-nucleon scattering
amplitudes, and one obtains an attractive isoscalar s-wave scattering length in
contradiction with experiments, where one finds a repulsive amplitude 
\cite{Mar81}. This discrepancy has been attributed to the existence of the 
$\Lambda(1405)$ which is located below the $K^- N$-threshold. From the 
analysis of $K^-p \rightarrow \Sigma \pi$ reactions it is known that this 
resonance couples strongly to the $I=0$ $K^-p$ state, and, therefore,
leads to repulsion in the $K^-p$ amplitude. 

\subsubsection*{\sc The Role of the $\Lambda(1405)$}  

Already in the sixties \cite{DWR67} there have been attempts to 
understand the $\Lambda(1405)$ as a bound state of the proton and the
$K^-$. In this picture, the underlying $K^-$-proton interaction is indeed
attractive as predicted by the chiral Lagrangians but the scattering 
amplitude is repulsive only because a bound state is formed. This concept is
familiar to the nuclear physicist from the deuteron, which is bound, because 
of the attractive interaction between proton and neutron. The existence of 
the deuteron then leads to a repulsive scattering length in spite of the 
attractive interaction between neutron and proton.  Chiral perturbation 
theory is based on a systematic expansion of the S-matrix elements in powers 
of momenta and, therefore, effects which are due to the proximity of a 
resonance, such as the $\Lambda(1405)$, will only show up in terms of rather 
high order in the chiral counting. Thus, it is not too surprising that the 
first two orders of the chiral expansion predict the wrong sign of the 
$K^-N$ amplitude. To circumvent this problem, the chiral perturbation 
calculation has been extended to either include an explicit $\Lambda(1405)$ 
state \cite{LBR94} or to use the interaction obtained from the leading order 
chiral Lagrangian as a kernel for a Lippman-Schwinger type calculation, 
which is then solved to generate a bound state $\Lambda(1405)$ \cite{KSW95}. 

\begin{figure}
\setlength{\epsfxsize=0.8}{\textwidth}
\centerline{\epsffile{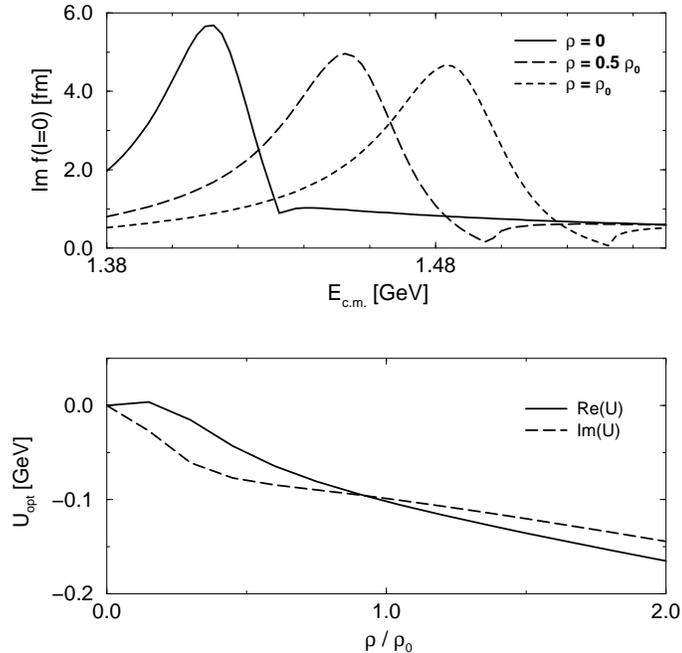}}
\caption{(a) Imaginary part of the $I=0$ $K^-$-proton scattering amplitude for
different densities. (b) real and Imaginary part of the $K^-$ optical
potential. }
\label{fig_lambda}
\end{figure}

This picture of the $\Lambda(1405)$ as a $K^-p$ bound state has recently
received some considerable interest in the context of in-medium corrections. 
In ref. \cite{Koc94b} it has been pointed out that in this picture  
as a result of the Pauli-blocking of the proton inside this bound state, 
the properties of the $\Lambda(1405)$ would be significantly changed in 
the nuclear environment. With increasing density, its mass increases and
the strength of the resonance is reduced (see Fig. \ref{fig_lambda}). 
Because of this shifting and `disappearance' of the $\Lambda(1405)$ in 
matter, the $K^-$ optical potential changes sign from repulsive to 
attractive at a density of about $1/4$ of nuclear matter density 
(see Fig. \ref{fig_lambda}) in agreement with a recent analysis of $K^-$
atoms \cite{FGB93}. These findings have been confirmed in ref. \cite{WKW96}.
This in-medium change of the $\Lambda(1405)$ due to the Pauli blocking can 
only occur if a large fraction of its wave function is indeed that of a 
$K^-$-proton bound state. Of course one could probably allow for a small 
admixture of a genuine three quark state without changing the results for
the measured kaon potentials. But certainly, if the $\Lambda(1405)$ is mostly 
a genuine three quark state, the Pauli blocking should not affect its 
properties. Therefore, it would be very interesting to confirm the mass 
shift of the $\Lambda(1405)$ for instance by a measurement of the missing 
mass spectrum of kaons in the reaction $p + \gamma \rightarrow \Lambda(1405) 
+ K^+$ at CEBAF. Thus, the atomic nucleus provides a unique laboratory to 
investigate the properties of elementary particles.  From our discussion it 
is clear that this in-medium change of the $\Lambda(1405)$ is not related 
to the restoration of chiral symmetry.

\subsubsection*{\sc Experimental results}

Phenomenologically, the attractive optical potential for the $K^-$ in nuclear
matter is of particular interest because it can lead to a possible kaon
condensation in neutron stars \cite{BLR94,BTK92}. This would limit the
maximum mass of neutron stars to about one and a half solar masses and give
rise to speculations about many small black holes in our galaxy \cite{BB94}.
Kaonic atoms, of course, only probe the very low density behavior of the kaon
optical potential and, therefore, an extrapolation to the large densities 
relevant for neutron stars is rather uncertain. Additional information
about the kaon optical potential can be obtained from heavy ion collision
experiments, where densities of more than twice nuclear matter density are
reached. 

Observables that are sensitive to the kaon mean-field potentials are
the subthreshold production \cite{FKL94a,FKL94b,LK95a} as well as kaon flow
\cite{LKL95,LK95b}. Given an attractive/repulsive mean-field potential
for the kaons it is clear that the subthreshold production is
enhanced/reduced. In case of the kaon flow an attractive interaction between
kaons and nucleons aligns the kaon flow with that of the nucleons whereas a
repulsion leads to an anti-alignment (anti-flow). However, both observables
are also extremely sensitive to the overall reaction dynamics, in particular 
to the properties of the nuclear mean field and to reabsorption processes
especially in case of the antikaons. Therefore, transport calculations 
are required in order to consistently incorporate all these effects. 

\begin{figure}[htp]
\setlength{\epsfxsize=\textwidth}
\centerline{\epsffile{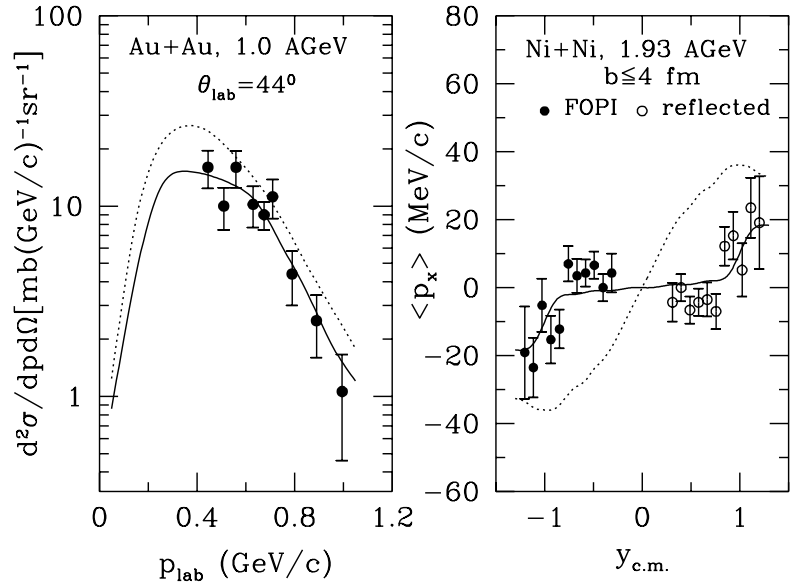}}
\caption{Kaon yield (left panel) and flow (right panel) in heavy ion 
collisions. The solid and dotted curves are results from transport model 
calculations with and without kaon mean-field potential, respectively. The data
are from refs. \protect\cite{Gro93,RFOPI95}.}
\label{fig_kaon_plus}
\end{figure}

In Fig. \ref{fig_kaon_plus} we show the result obtained from such  
calculations for the $K^+$ subthreshold production and flow together with
experimental data (solid circles) from the KaoS collaboration \cite{Gro93} 
and from the FOPI collaboration \cite{RFOPI95} at GSI.
Results are shown for calculations without any 
mean fields (dotted curves) as well as with a repulsive mean field obtained
from the chiral Lagrangian (solid curves), which is consistent with a simple 
impulse approximation. The experimental data are nicely reproduced when 
the kaon mean-field potential is included.
Although the subthreshold production of kaons can also be explained without 
any kaon mean field \cite{CHK96,MCM94}, the assumption used in these
calculations that a lambda particle has the same mean-field potential
as a nucleon is not consistent with the phenomenology of hypernuclei
\cite{GH95}. Since it is undisputed that these observables are sensitive 
to the in-medium kaon potential, a systematic investigation including 
all observables should eventually reveal more accurately the strength 
of the kaon potential in dense matter. Additional evidence for a repulsive 
$K^+$ potential comes from $K^+$-nucleus experiments. There the measured 
cross sections and angular distributions can be pretty well understood 
within a simple impulse approximation. Actually it seems that an additional 
(15\%) repulsion is required to obtain an optimal fit to the data 
\cite{CE92}, which has been suggested as a possible evidence for a swelling 
of the nucleon size or a lowering of the omega meson mass in the nuclear 
medium \cite{BDS88}.  

\begin{figure}[htp]
\setlength{\epsfxsize=\textwidth}
\centerline{\epsffile{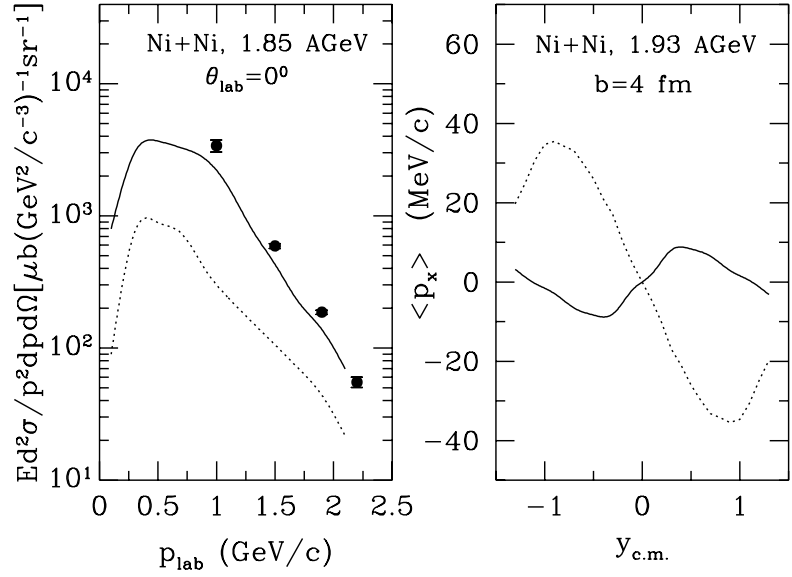}}
\caption{Antikaon yield (left panel) and flow (right panel) in heavy ion 
collisions. The solid and dotted curves are results from transport model 
calculations with and without kaon mean-field potential, respectively. The data
are from ref. \protect\cite{Schr94}.}
\label{fig_kaon_minus}
\end{figure}

As for the $K^-$, both chiral perturbation theory and dynamical models of 
the $K^-$-nucleon interaction \cite{Koc94b} indicate the existence of an 
attractive mean field, so the same observables can be used \cite{LKF94,LK96}.
However, the measurement and interpretation of $K^-$ observables is more 
difficult since it is produced less abundantly. Also reabsorption effects 
due to the reaction $K^-  N \rightarrow \Lambda \pi$ are strong, which 
further complicates the analysis. Nevertheless, the effect of the attractive 
mean field has been shown to be significant as illustrated in Fig. 
\ref{fig_kaon_minus}, where results from transport calculations with 
(solid curves) or without (dotted curves) attractive mean field for the 
antikaons \cite{LKF94} are shown. It is seen that the data (solid circles) 
on subthreshold $K^-$ production \cite{Schr94} support the existence 
of an attractive antikaon mean-field potential. For the $K^-$ flow,
there only exist very preliminary data from the FOPI collaboration 
\cite{Rit96}, which seem to show that the $K^-$'s have a positive flow 
rather an antiflow, thus again consistent with an attractive antikaon 
mean-field potential. 

Let us conclude this section on kaons by pointing out that the
properties of kaons in matter are qualitatively described in chiral 
perturbation
theory. Although some of the effects comes from the Weinberg-Tomozawa 
vector type interaction, which contributes because contrary to the pion the 
kaon has only one light quark, in order to explain the experimental data on
kaon subthreshold production and flow in heavy ion collisions 
an additional attractive scalar interaction is required, 
which is related to the 
explicit breaking of chiral symmetry and higher order corrections in the chiral
expansion. 

\subsection{\it Etas }

The properties of etas are not only determined by consideration of chiral
symmetry but also by the explicit breaking of the $U_A(1)$ axial symmetry due
to the axial anomaly in QCD. As a result, the singlet eta, which would be a
Goldstone boson if $U_A(1)$ were not explicitly broken, becomes heavy. 
Furthermore, because of SU(3) symmetry breaking -- the strange quark
mass is considerably heavier than that of up and down quark -- the octet
eta and singlet eta mix, leading to the observed particles $\eta$ and $\eta'$. 
The mixing is such that the $\eta'$ is mostly singlet and thus heavy, and the
$\eta$ is mostly octet and therefore has roughly the mass of the kaons.
If, as has been speculated \cite{Shu94,Sch96}, the $U_A(1)$ symmetry is 
restored at high temperature due to the instanton effects, one would expect 
considerable reduction in the masses of the etas as well as in their mixing
\cite{KKM96,HW96}.  A dropping eta meson in-medium mass is expected
to provide a possible explanation for the observed enhancement of low 
transverse momentum etas in SIS heavy ion experiments at subthreshold 
energies \cite{Ber94}. However, an analysis of photon spectra 
from heavy ion collisions at SPS-energies puts an upper limit on the 
$\eta/\pi^0$ ratio to be not more than 20 \% larger for central than for
peripheral collisions \cite{Dre96}. This imposes severe constraints on the
changes of the $\eta$ properties in hadronic matter. 

We note that the restoration of chiral symmetry is important in the 
$\eta - \eta'$ sector, because without the $U_A(1)$ breaking both would be 
Goldstone bosons of an extended $U(3) \times U(3)$ symmetry. 
However, it is still being debated what the effects precisely are.

\section{VECTOR MESONS} 

Of all particles it is
probably the $\rho$-meson which has received the most attention in regards of
in-medium corrections. This is mainly due to the fact that the $\rho$ is
directly observable in the dilepton invariant mass spectrum. Also, since the
$\rho$ carries the quantum numbers of the conserved vector current, its
properties are related to chiral symmetry and can, as we shall discuss, be
investigated using effective chiral models as well as current algebra and QCD
sum rules.  That possible changes of the $\rho$ can be observed in the 
dilepton spectrum measured in heavy ion collisions has been first 
demonstrated in ref. \cite{Koc92} and then studied in more detail in 
\cite{WCM93,LK95c}. The in-medium properties of the $a_1$ are closely 
related to that of the $\rho$ since they are chiral partners and their 
mass difference in vacuum is due to the spontaneous breaking of chiral 
symmetry \cite{Wei67}. Unfortunately, there is no direct method to measure 
the changes of the $a_1$ in hadronic matter.  The $\omega$-meson on the 
other hand is a chiral singlet, and the relation of its properties 
to chiral symmetry is thus not so direct. However, calculations 
based on QCD sum rules predict also changes in the $\omega$ mass as one 
approaches chiral restoration. Observationally, the $\omega$ can probably 
be studied best due to its rather small width and its decay channel into 
dileptons. Finally, there is the $\phi$ meson. In an extended SU(3) chiral 
symmetry, the $\phi$ and the $\omega$ are a superposition of the singlet 
and octet states with nearly perfect mixing, i.e. the $\phi$ contains only 
strange quarks whereas the $\omega$ is made only out of light quarks. 
Both QCD sum rules \cite{AK94a} and effective chiral models
\cite{Son96} predict a lowering of the $\phi$-meson mass in medium.

The question on whether or not masses of light hadrons change in the medium
has received considerable interest as a 
result of the conjecture of Brown and Rho
\cite{BR91}, which asserts that the masses of all mesons, with the exception 
of the Goldstone Bosons, should scale with the quark condensate. While the 
detailed theoretical foundations of this conjecture are still being worked on
\cite{AB93,FM96,BBR96}
the basic argument of Brown and Rho is as follows. Hadron masses, such as that
of the $\rho$-meson,  violate scale
invariance, which is a symmetry of the classical QCD Lagrangian. In QCD scale
invariance is broken on the quantum level by the so-called trace anomaly (see
e.g. \cite{DGH92}), which is proportional to the Gluon condensate. Thus one
could imagine that with the disappearance of the gluon 
condensate, i.e. the bag
pressure, scale invariance is restored, which on the hadronic level implies
that hadron masses have to vanish. Therefore, one could argue that hadron
masses should scale with the bag-pressure as originally proposed by Pisarski
\cite{Pis82}. But the conjecture of Brown and Rho goes even further. They
assume that the gluon condensate can be separated into a hard and soft part,
the latter of which scales with the quark condensate and is also responsible
for the masses of the light hadrons. This picture finds some support from 
lattice
QCD calculations in that the gluons condensate drops by about 50\% at the
chiral phase transition \cite{KB93b}. To what extent this is also reflected
in changes in hadron masses in not clear at the moment (see section
\ref{lattice}), although one should mention that the rise in the entropy
density close to the critical temperature can be explained if one assumes the
hadron masses to scale with the quark condensate \cite{KB93b}. 
Another aspect of the Brown and Rho scaling is that once the scaling hadron
masses are introduced in the chiral Lagrangian, only tree-level diagrams are
needed as the contribution from higher order diagrams is expected to 
be suppressed. In their picture, Goldstone bosons
are not subject to this scaling, since they receive their mass from the
explicit chiral symmetry breaking due to finite current quark masses, which are
presumably generated at a much higher (Higgs) scale. 

\subsection{\it The rho meson}
\label{rho}
Since the $\rho$ is a vector meson, it couples directly to the isovector
current which then results in the direct decay of the $\rho$ into virtual
photons, i.e. dileptons. Consequently, properties of the $\rho$ meson can be
investigated by studying two-point correlation functions of the the isovector
currents, i.e.,
\be   
\Pi_{\mu\nu}(q) = i\int e^{iqx} \langle T J_{\mu}(x)J_{\nu}(0) 
\rangle_\rho d^4x.
\ee
The masses of the rho meson and its excitations
($\rho' \ldots$) correspond to the positions of the poles of this correlation 
function. This is best seen if one assumes that the current field identity 
\cite{Sak69} holds, namely that the current operator is proportional to the 
$\rho$-meson field. In this case the above correlation function is identical, 
up to a constant, to the $\rho$-meson propagator. The imaginary part of this 
correlation function is also directly proportional to the electron-positron 
annihilation cross section \cite{GK91}, 
where the $\rho$-meson is nicely seen. In 
addition, at higher center-of-mass energies, one sees a continuum in the 
electron-positron annihilation cross section, which corresponds 
to the excited states of the rho mesons as well as to the onset of 
perturbative QCD processes. 

In-medium changes of the $\rho$ meson can be addressed theoretically by 
evaluating the current-current correlator in the hadronic environment. 
The current-current correlator can be evaluated either in effective chiral 
models, or using current algebra arguments, or directly in QCD. In the 
latter, one evaluates the correlator in the deeply Euclidean region
($q^2 \rightarrow - \infty$) using the Wilson expansion, where all the long
distance physics is expressed in terms of vacuum expectation values of quark
and gluon operators, the so-called condensates. Dispersion relations are then
used to relate the correlator in the Euclidean region to that for positive
$q^2$, where the hadronic eigenstates are located. One then assumes a certain
shape for the phenomenological spectral functions, typically a delta function,
which represents the bound state, and a continuum, which represents the
perturbative regime. These so called QCD sum rules, therefore, relate the
observable hadronic spectrum with the QCD vacuum condensates (for a review of
the QCD sum-rule techniques see e.g. \cite{RRY85}). These relations can then
either be used to determine the condensates from measured hadronic spectra,
or, to make predictions about changes of the hadronic spectrum due to in-medium
changes of the condensates. 

\begin{figure}[htp]
\setlength{\epsfxsize}{\textwidth}
\centerline{\epsffile{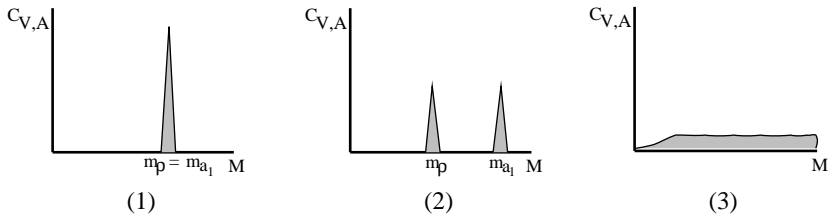}}
\caption{Several possibilities for the vector and axial-vector spectral
functions in the chirally restored phase.}
\label{fig_corr}
\end{figure}

Similarly, one can study the properties of $a_1$ meson in the nuclear medium 
through the axial vector correlation function.  Then, 
once chiral symmetry is restored, there should be no observable difference
between left-handed and right-handed or equivalently vector and axial vector
currents. Consequently, the vector and axial vector correlators should be
identical. Often, this identity of the correlators is identified with the
degeneracy of the $\rho$ and $a_1$ mesons in a chirally symmetric world. This,
however, is not the only possibility, as was pointed out by Kapusta and Shuryak
\cite{KS94}. There are at least three qualitatively different scenarios, for
which the vector and axial vector correlator are identical (see Fig. 
\ref{fig_corr}). 
\begin{enumerate}
\item
The masses of $\rho$ and $a_1$ are the same. The value of the common mass,
however, does not follow from chiral symmetry arguments alone.
\item
The mixing  of the spectral functions, i.e, both the vector and axial-vector 
spectral functions have peaks of similar strength at both the mass of the 
$\rho$ and the mass of the $a_1$.  
\item
Both spectral functions could be smeared over the entire mass range. Because 
of thermal broadening of the mesons and the onset of deconfinement, the 
structure of the spectral function may be washed out, and it becomes 
meaningless to talk about mesonic states.
\end{enumerate}

\subsubsection*{\sc Finite temperature}

At low temperatures, where the heat bath consists of pions only,
one can employ  current algebra as well as PCAC to obtain an essentially 
model-independent result for the properties of the $\rho$.
Using this technique Dey,
Eletzky and Ioffe \cite{DEI90} could show that to leading order in the
temperature, $T^2$, the mass of the $\rho$-meson does not change. 
Instead one finds an admixture of the axial-vector correlator, i.e. that
governed by the $a_1$-meson. Specifically to this order the vector correlator 
is given by
\be
C_V(q,T) = (1 - \epsilon) \, C_V(q, T= 0 ) + \epsilon \, C_A(q,T=0) 
+ {\cal O}(T^4),
\label{mix_low_T}
\ee
with $\epsilon = T^2/(6 f_\pi^2)$. Here $C_V, \, C_A$ stand for the
vector-isovector and axial-vector-isovector correlation functions,
respectively. One should note, that to the same order the chiral condensate is
reduced \cite{GL89},
\be
\frac{<\bar{q}q >_T}{<\bar{q}q>_0} = 1 - \frac{T^2}{8 f_\pi^2} + {\cal O}(T^4).
\ee
Therefore, to leading order a drop in the chiral condensate does not affect
the mass of the $\rho$-meson but rather reduces its coupling to the vector
current and in particular induces an admixture of the $a_1$-meson. 
This finding is at variance with the Brown-Rho scaling hypothesis. The reason
for this difference is not yet understood.
The admixture of the axial correlator is directly related with the onset of
chiral restoration. If chiral symmetry is restored, the vector and
axial-vector correlators should be identical. The result of Dey et al. suggests
that this is achieved by a mixing of the two instead of a degeneracy of
the $\rho$ and $a_1$ masses.  If the mixing is complete, i.e. $\epsilon 
= 1/2$, then the extrapolation of the low temperature result \raf{mix_low_T} 
would give a critical temperature of $T_c = \sqrt{3} f_\pi \simeq 164 \, MeV$, 
which is surprisingly close to the value given by recent lattice calculations.

Corrections to the order $T^4$ involve physics beyond chiral symmetry. 
As nicely discussed in ref. \cite{EI95}, to this order the contributions 
can be separated into two distinct contributions.  The first arises if a 
pion from the heat bath couples via a derivative coupling to the current 
under consideration. In this case 
the contribution is proportional to the invariant 
pion density 
\be
n_\pi = \int \frac{d^3 q}{2 \omega (2 \pi)^3} {\rm e}^{- \omega / T} \sim T^2,
\ee
and the square of the pion momentum $q^2 \sim T^2$. A typical example would 
be for instance the self-energy correction to the $\rho$ meson from the 
standard two-pion loop diagram involving the p-wave $\pi \pi \rho$ coupling.
The other contribution comes from interactions of two pions from the heat bath
with the current, without derivative couplings. This is proportional to the
square of the invariant pion density and thus $\sim T^4$ as well. 
These pure density contributions again can be evaluated in a model
independent fashion using current algebra techniques and, as before, do not
change the mass of the $\rho$-meson, but change the coupling to the current 
and induce a mixing with the axial-vector correlator. Actually, it can be shown
\cite{EI95} that to all orders in the pion density these pure density effects
do not change the $\rho$ mass but induce mixing and reduce the coupling. At the
same time, however, these contributions reduce the chiral condensate. 

The contributions due to the finite pion momenta have been estimated in
\cite{EI95} to give a downward shift of both the $\rho$ and $a_1$ mass of 
about 
\be
\frac{\delta m}{m} \simeq 10 \%.
\ee
for temperatures of $\sim 150 - 200 \, \rm MeV$. We should point out, 
however, that in this analysis those changes in the masses arise from 
Lorentz-nonscalar condensates in the operator product expansion. Therefore, 
they are not directly related to the change of the Lorentz-scalar 
chiral-condensate. 

Effective chiral models also have been used to explore 
the order $T^4$ corrections. Ref. \cite{Pis95} reports that the mass of 
the $\rho$ drops whereas that of the $a_1$ increases to this order. 
This is somewhat at variance with the findings of \cite{Son93} where
no drop in the mass of the $\rho$ but a decrease in the $a_1$ mass
has been found. In this calculation, however, the terms leading to the 
order $T^4$
changes have not been explicitly identified but rather a calculation to one
loop order has been carried out. Presently, the difference in these results is
not understood. Also the difference between the analysis of \cite{EI95} and
the effective chiral models is not resolved yet. The analysis of \cite{EI95} 
uses dispersion relations to relate phenomenological space-like photon-pion 
amplitudes with the time-like ones needed to calculate the mass shift. The 
effective Lagrangian methods, on the other hand, rely on pion-pion 
scattering data as well as measured decay width in order to fix their model 
parameters. One would think that both methods should give a reasonably 
handle on the leading order momentum-dependent couplings.

At higher temperatures as well as at finite density vacuum 
effects due to the virtual pair correction or the nucleon-antinucleon 
polarization could become important and they tend to reduce vector meson 
masses \cite{SXK95}.  Because of the large mass of the nucleon, these 
corrections however, do not affect the leading temperature result 
$\sim T^2$ discussed previously. Also this approach assumes that the
physical vacuum consists of nucleon-antinucleon rather than quark-antiquark 
fluctuations. Whether this is the correct picture is, however, not yet 
resolved.

\subsubsection*{\sc Finite density}

Since the density effect on the chiral condensate is much stronger 
than that of the temperature, one expects the same for the rho meson mass.
But the situation is more complicated at finite density as
one cannot make use
of current algebra arguments and thus model-independent result as the one 
discussed previously are not available at this time. Present model 
calculations, however, disagree even on the sign of the mass shift. One 
class of models \cite{RW93,CS93,HFN93,AKL92,KW96} considers the $\rho$ as a 
pion-pion resonance and calculates its in-medium modifications due to those 
for the pions as discussed in section \ref{pion}. 
These calculations typically show an increased width of the rho since it now 
can also decay into pion-nucleon-hole or pion-delta-hole states. At the same
time this leads to
an increased strength below the rho meson mass. 
The mass of the rho meson, defined as the position where the real part of the
correlation function goes through zero, is shifted only very little. Most
calculations give an upward shift but also a small downward shift has been 
reported \cite{KW96,KW97}. This difference seems to  depend on 
the specific choice of the cutoff functions for the vertices involved 
\cite{KW97b}, and 
thus is  model dependent. However, the imaginary part of the correlation
function, the relevant quantity for the dilepton measurements, hardly depends 
on a small upward or downward shift of the rho. The important and apparently 
model independent feature is the increased strength at low invariant masses 
due to the additional decay-channels available in nuclear matter.

Calculation using QCD-sum rules \cite{HL92} predict a rather strong decrease 
of the $\rho$-mass with density ($\sim 20\%$ at nuclear matter density),
which is similar to the much discussed Brown-Rho scaling \cite{BR91}.
Here, as in the finite temperature case, the driving term is the four 
quark condensate which is assumed to factorize
\be
\langle (\bar{q}\gamma_{\mu}\lambda^{a}q)
(\bar{q}\gamma^{\mu}\lambda^{a}q)\rangle _\rho&\approx& -\frac{16}{9}
\langle \bar{q} q \rangle_\rho^2.
\label{fac}
\end{eqnarray}
To which extent this factorization is correct at finite density is not
clear. Also, when it comes to the parameterization of the phenomenological
spectral distribution, these calculations usually assume the standard pole
plus continuum form with the addition of a so called Landau damping
contribution at $q^2 = 0$ \cite{HL92}. The additional strength below the mass
of the rho as predicted by the 
previously discussed models is usually
ignored. An attempt, however, has been made to see if a spectral distribution
obtained from the effective models described above does saturate the QCD 
sum rules \cite{AK93}. These authors found that they could only saturate the 
sum rule if they assumed an additional mass shift of the rho-meson peak 
downwards by $\sim 140$ MeV. However, in this calculation 
only leading order density corrections to the quark condensates whereas 
infinity order density effects have been 
taken into account in order to calculate the spectrum. Another 
comparison with the QCD sum rules has recently been carried out in
ref. \cite{KW97} where a reasonable saturation of the sum-rule is reported
using a similar model for the in medium correlation function.

There are also attempts to use photon-nucleon data in order to estimate
possible mass shifts of the $\rho$ and $\omega$ mesons. In ref. \cite{FS96} a
simple pion and sigma exchange model is used in order to fit data for
photoproduction of $\rho$- and $\omega$-mesons. Assuming vector dominance, this
model is then used to calculate the self-energy of these vector mesons in
nuclear matter. The authors find a downward shift of the $\rho$ of about $18
\%$ at nuclear matter density, in rough agreement with the prediction from QCD
sum rules. Quite to the contrary, ref. \cite{EI96} using photoabsorption data
and dispersion relations find an upwards shift of 10~MeV or 50~MeV for the 
longitudinal and transverse part of the $\rho$, respectively. This result, 
however, is derived for a $\rho$-meson which is not at rest in the nuclear 
matter frame, and, therefore a direct comparison of the two predictions 
is not possible.

Very recently Friman and Pirner have pointed out that
the $\rho$-meson couples very
strongly with the $N^*(1720)$ resonance \cite{QM96,FP97}. The coupling is of
p-wave nature and, similarly to the pion coupling to delta-hole states, the
$\rho$ may mix with $N^*$-nucleon-hole states resulting in a modified
dispersion relation for the $\rho$-meson in nuclear matter. Since the coupling
is p-wave, only $\rho$-mesons with finite momentum are modified and shifted to
lower masses. This momentum dependence of the low mass enhancement in the
dilepton spectrum is an unique prediction which can be tested in experiment.

\subsubsection*{\sc Experimental results}

First measurement of dileptons in heavy ion collisions have been
carried out by the DLS collaboration \cite{DLS88a,DLS88b} at the BEVALAC. 
The first published data based on a limited data set could be well 
reproduced using state of the art transport models \cite{WCM93,XWK90,WBC90}
without any additional in-medium corrections. However, a recent reanalysis
\cite{DLS97}
including the full data set seems to show a considerable increase over the
originally published data. It remains to be seen if these new data can also be
understood without any in-medium corrections to hadronic properties.
\begin{figure}[htp]
\setlength{\epsfxsize=1.0}{\textwidth}
\centerline{\epsffile{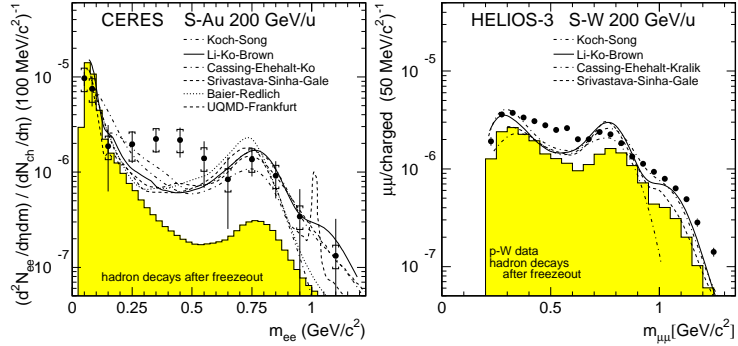}}
\caption{Dilepton invariant mass spectrum from S+Au and S+W collisions
at 200 GeV/nucleon without dropping vector meson masses.
The experimental statistical errors are shown as bars and the systematic 
errors are marked independently by brackets.
(The figure is from ref. \protect\cite{Dre97})}
\label{drees1}
\end{figure}
Recently, low mass dilepton data taken at the CERN SPS have been published 
by the CERES collaboration \cite{CERES95}.  Their measurements for p+Be and 
p+Au could be well understood within a hadronic cocktail, which takes into 
account the measured particle yields from p+p experiments and their decay 
channels into dileptons. In case of the heavy ion collision, S+Pb, however, 
the hadronic cocktail considerably underpredicted the measured data in 
particular in the invariant mass region of $300 \, \rm MeV \leq M_{inv} 
\leq 500 \, \rm MeV$. A similar enhancement 
has also been reported by the HELIOS-3 collaboration \cite{HELIOS95}.  
\begin{figure}[htp]
\setlength{\epsfxsize=1.0}{\textwidth}
\centerline{\epsffile{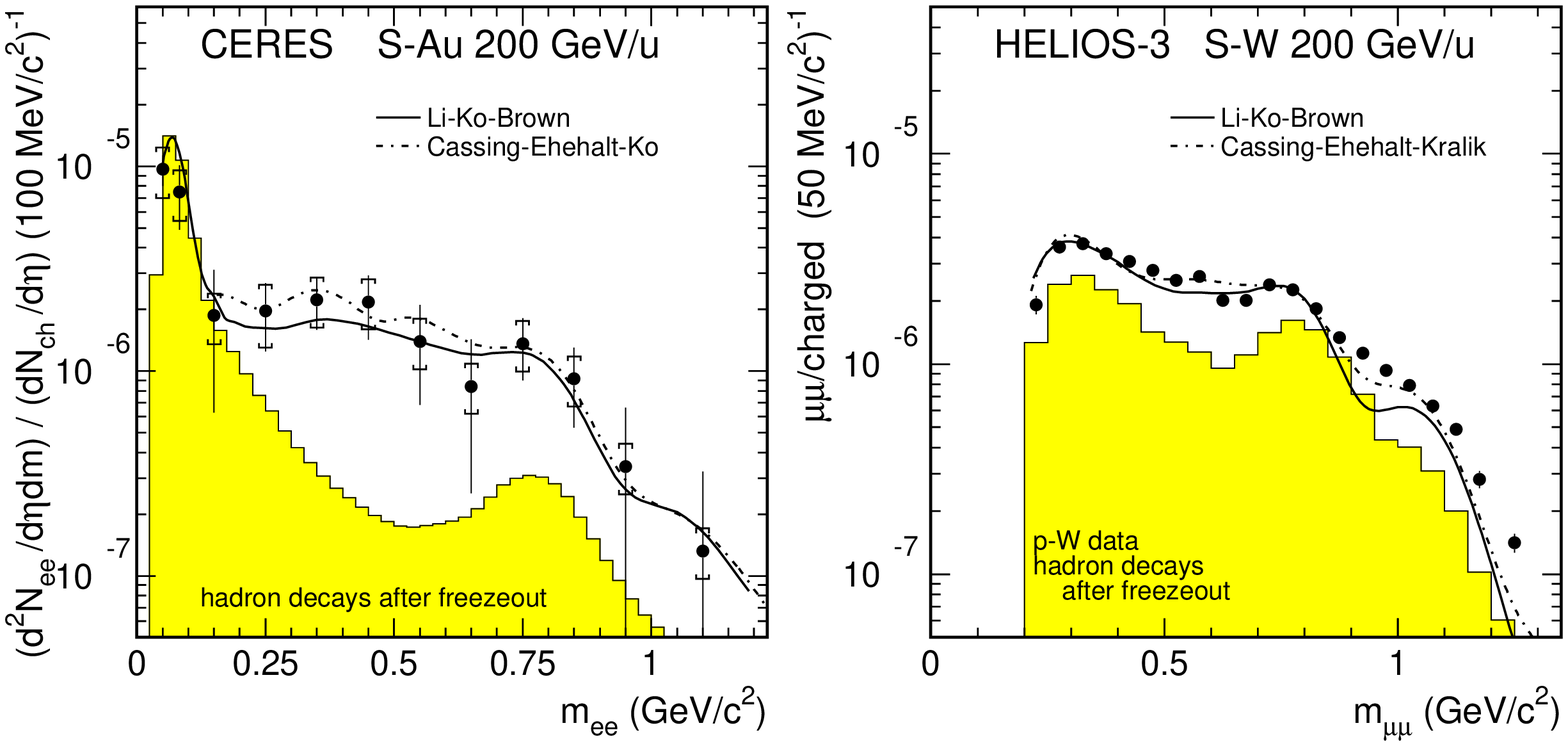}}
\caption{Same as Fig. \protect\ref{drees1} with dropping vector meson masses.
(The figure is from ref. \protect\cite{Dre97})}
\label{drees2}
\end{figure}

Of course it is well-known 
that in a heavy ion collisions at SPS-energies a hadronic fireball consisting 
predominantly of pions is created. These pions can pairwise annihilate into 
dileptons giving rise to an additional source, which has not been included 
into the hadronic cocktail. While the pion annihilation contributes to 
the desired mass range, many detailed calculations \cite{KS96,CEK95,SSG96},
which have taken this channel into account, still underestimate the data by 
about a factor of three as shown Fig. \ref{drees1}. 
Also, additional hadronic processes, such as $\pi\rho\to\pi e^+e^-$, 
have been taken into account \cite{Hag96,BDR96}. However, 
`conservative' calculations could at best reach the lower end of the sum 
of statistical and systematic error bars of the CERES data. 
Furthermore, in-medium corrections to the pion annihilation 
process together with corrections to the $\rho$-meson spectral distribution 
have been considered using effective chiral models \cite{SKL96,RCW95}. But, 
these in-medium corrections, while enhancing the contribution of the pion 
annihilation channel somewhat, are too small to reproduce the central 
data points. Only models which allow for a dropping of the $\rho$-meson 
mass give enough yield in the low mass region \cite{CEK95,LKB95,LKB96,LKBS96}
as shown in Fig. \ref{drees2}. In these models, the change of the rho meson 
mass is obtained from either the Brown-Rho scaling \cite{LKB95,LKB96,LKBS96} 
or the QCD sum rules \cite{CEK95}. Furthermore, the vector dominance is 
assumed to be suppressed in the sense that the pion electromagetic from 
factor, which in free space is dominated by the rho meson and is proportional
to the square of the rho meson mass, is reduced as a result of the dropping
rho meson mass. At finite temperature, such a suppression has been 
shown to exist in both the hidden gauge theory \cite{SLK95} and 
the perturbtive QCD \cite{DS94}, where the pion electromagnetic form factor 
is found to decrease at the order $T^2$. This effect is 
related to the mixing between the vector-isovector and axial-vector-isovector
correlators at finite temperature we discussed earlier. In most studies 
of dilepton production without a dropping rho meson mass, this effect has 
not been included.  Although the temperature reached in heavy 
ion collisions at SPS energies is high and the ratio of pions to baryons 
in the final state is about 5 to 1, the authors in 
\cite{CEK95,LKB95,LKB96,LKBS96} found that they had to rely on the nuclear 
density rather than on the temperature in order to reproduce the data. 

There are also preliminary data from the Pb+Au collisions at 150 
GeV/nucleon \cite{Dre97}, which also seem to be consistent with the dropping
in-medium rho meson scenario as well. Unfortunately, the experimental errors 
in this case are even larger than in the S+Au data and thus these new,
preliminary data do not further discriminate between the dropping rho-mass and
more conventional scenarios.  
Thus, additional data with reduced error bars are needed before firm 
conclusions about in-medium changes of the rho meson mass can be drawn. 
Although the data from the HELIOS-3 collaboration shown in the right
panels of Figs. \ref{drees1} and \ref{drees2} do have very small errors,
these are only the statistical ones while the systematic errors are not
known. 

To summarize this section on the rho meson,
our present theoretical understanding for possible in-medium
changes at low temperatures and vanishing baryon density indicates that
one expects no (to leading order) and a small (to next to leading order) 
changes of the $\rho$-mass. At the same time to leading order in the 
temperature, the chiral condensate is reduced. Therefore, a direct connection 
between changes in the chiral condensate and the mass of the $\rho$ meson 
has not been established as assumed in the Brown-Rho scaling. 
However, it may well be that the latter is only valid at higher temperature
near the chiral phase transition. 
At finite density additional effects such as nuclear many-body excitations 
come into play, which make the situation much more complicated. 
Nevertheless, the observed enhancement of low mass dileptons from
CERN-SPS heavy ion collisions can best be described by a dropping 
rho meson mass in dense matter.
Finite density and zero temperature is another area where possible 
changes of the vector mesons can be measured in experiment.  Photon, 
proton and pion induced dilepton production from nuclei will soon be measured 
at CEBAF \cite{Pre95} and at GSI \cite{str95}. In principle these measurements 
should be able to determine the entire spectral distribution. If the 
predictions of the Brown-Rho scaling and the QCD sum rules are correct, 
mass shifts of the order of 
100 MeV should occur, which would be visible in these experiments. 
Furthermore, by choosing appropriate kinematics, the properties of the 
$\rho$ meson at rest as well as at finite momentum with respect to the 
nuclear rest frame can be measured. 

\subsection{\it The omega meson}

The omega meson couples to the isoscalar part of
the electromagnetic current. In QCD sum rules it is also dominated
by the four quark condensate. If the latter is reduced in medium, then the 
omega meson mass also decreases. Assuming factorization for the four
quark condensate, Hatsuda and Lee \cite{HL92} showed that the change of 
the omega meson mass in dense nuclear matter would be similar to that of 
the rho 
meson mass. The dropping omega meson mass in medium can also be obtained
by considering the nucleon-antinucleon polarization in medium 
\cite{SXK95,JPW94,SH94} as in the case of the rho meson. 
On the other hand, using effective chiral Lagrangians, one finds 
at finite temperature an even smaller change in the mass of the omega 
as compared to that of the rho \cite{Son93}.
Because of its small decay width, which is about an 
order of magnitude smaller than that of rho meson, most dileptons
from omega decay in heavy ion collisions are emitted after freeze out,
where the medium effects are negligible. However, with appropriate
kinematics an omega can be produced at rest in nuclei in reactions
induced by the photon, proton and pion. By measuring the decay of the 
omega into dileptons allows one to determine its properties in the
nuclear matter. Such experiments will be carried out at CEBAF \cite{Pre95} 
and GSI \cite{str95}.

\subsection{\it The phi meson}

For the phi meson, the situation is less ambiguous than the rho meson 
as both effective chiral Lagrangian and QCD sum-rule studies predict 
that its mass decreases in medium. Based on the hidden gauge theory, 
Song \cite{Son96} finds that at a temperature of $T=200$ MeV the 
phi meson mass is reduced by about 20 MeV. 
The main contribution is from the thermal kaon 
loop. In QCD sum rules, the reduction is much more appreciable, i.e., about
200 MeV at the same temperature \cite{AK94a}. The latter is due to 
the significant decrease of the strange quark condensate at finite 
temperature as a result of the abundant strange hadrons in hot matter.
Because of the relative large strange quark mass compared to the up and
down quark masses, the phi meson mass in QCD sum rules is mainly 
determined by the strange quark condensate instead of the four quark 
condensate as in the case of rho meson mass. However, the temperature
dependence of the strange quark condensate in \cite{AK94a} is determined
from a non-interacting gas model, so effects due to interactions, which
may be important at high temperatures, are not included. In \cite{Son96},
only the lowest kaon loop has been included, so the change of kaon properties 
in medium is neglected. As shown in \cite{KLQ92}, this would reduce the
phi meson mass if the kaon mass becomes small in medium. Also, vacuum 
effects in medium due to lambda-antilambda polarization has also been 
shown to reduce the phi meson in-medium mass \cite{KH95}.

QCD sum rules have also been used in studying phi meson mass at finite 
density \cite{HL92}, and it is found to decrease by about 25 MeV at
normal nuclear matter density. 

Current experimental data on phi meson mass from measuring the $K\bar K$ 
invariance mass spectra in heavy ion collisions at AGS energies do not
show a change of the phi meson mass \cite{Aki96}. This is not surprising 
as these kaon-antikaon pairs are from the decay of phi mesons at freeze
out when their properties are the same as in free space. If a phi meson
decays in medium, the resulting kaon and antikaon would interact with 
nucleons, so their invariant mass is modified and can no longer be 
used to reconstruct the phi meson. However, future experiments on 
measuring dileptons from photon-nucleus reactions at CEBAF \cite{Pre95} 
and heavy ion collisions at GSI \cite{str95} will provide useful
information on the phi meson properties in dense nuclear matter. For heavy ion
collisions at RHIC energies, matter with low-baryon chemical potential 
is expected to be formed in central rapidities. Based on the QCD-sum rule
prediction for the mass shift of the phi meson it has been suggested
that a low mass phi peak at $\sim 880$ MeV besides the normal one at
1.02 GeV appears in the dilepton spectrum if a first-order phase transition
or a slow cross-over between the quark-gluon plasma and the hadronic matter 
occurs in the collisions \cite{AK94b,AK94c}.  The low-mass phi peak is 
due to the nonnegligible duration time for the system to stay near the 
transition temperature compared with the lifetime of a phi meson in 
vacuum, so the contribution to dileptons from phi meson decays in the
mixed phase becomes comparable to that from their decays at freeze out. 
Without the formation of the quark-gluon plasma, the low-mass phi peak
is reduced to a shoulder in the dilepton spectrum.  Thus, one can 
use this double phi peaks in the dilepton spectrum as a signature for 
identifying the quark-gluon plasma to hadronic matter phase transition 
in ultrarelativistic heavy ion collisions.
\section{BARYONS}

As far as the in medium properties of the baryons are concerned most
is known about the nucleon. But also the properties of the  hyperons
such as the $\Lambda$ can be determined in the medium by studying hypernuclei.
Also the $\Lambda(1405)$, which we have discussed in connection with the $K^-$
optical potential, is another example for in-medium effects of baryons. In 
the following we will limit ourself on a brief discussion of how the nucleon
properties can be viewed in the context of chiral symmetry.

As mentioned briefly in the introduction, medium effects on a nucleon
due to many-body interactions have long been studied, leading to an 
effective mass, which is generally reduced as a result of the finite 
range of the nucleon interaction and higher order effects. Its relation 
to chiral symmetry restoration is best seen through the QCD sum rules
\cite{DL90,CFG92,CFG91,JCF93,JCF94,HHP90,AB90}. Using in-medium 
condensates, it has been shown that the change of scalar quark 
condensate in medium leads to an attractive scalar potential which
reduces the nucleon mass, while the change of vector quark condensate
leads to a repulsive vector potential which shifts its energy. 
The nucleon scalar and vector self-energies in this study are given, 
respectively, by
\begin{eqnarray}\label{cohen}
\Sigma_S&\approx& -\frac{8\pi^2}{M_B^2}(\langle\bar qq\rangle_\rho-
\langle\bar qq\rangle_0)
\approx -\frac{8\pi^2}{M_B^2}\frac{\Sigma_{\pi N}}{m_u+m_d}
\rho_N,\nonumber\\
\Sigma_V&\approx& \frac{64\pi^2}{3M_B^2}\langle q^+q\rangle_\rho=
\frac{32\pi^2}{M_B^2}\rho_N,
\end{eqnarray}
where the Borel mass $M_B$ is an arbitrary parameter. With $M_B\approx m_N$, 
and $m_u+m_d\approx 11$ MeV, these self-energies have magnitude of a
few hundred MeV at normal nuclear density. These values are similar to 
those determined from both the Walecka model \cite{SW86} and the 
Dirac-Brueckner-Hartree-Fock (DBHF) approach \cite{Mac89} based on the 
meson-exchange nucleon-nucleon interaction.
Experimental evidences for these strong scalar and vector potentials
have been inferred from the proton-nucleus scattering at intermediate 
energies \cite{RHC92} via the Dirac phenomenology 
\cite{ACM79,KCJ85,HCC90,CHC93} in which the Dirac equation with scalar 
and vector potentials is solved.  

\begin{figure}[htp]
\setlength{\epsfxsize=\textwidth}
\centerline{\epsffile{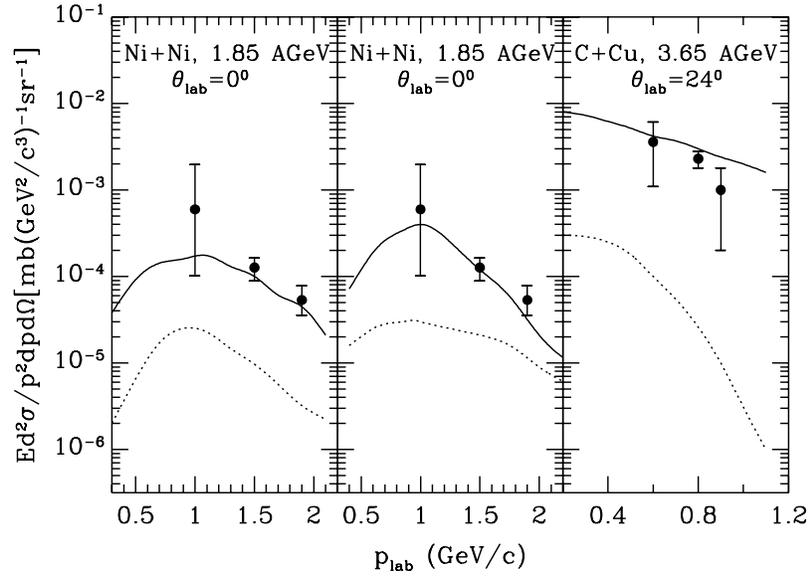}}
\caption{Antiproton momentum spectra from Ni+Ni collisions 
at 1.85 GeV/nucleon, and C+Cu collisions at 3.85 GeV/nucleon.
The left, middle, and right panels are from Refs. \protect\cite{antip},
\protect\cite{MOS94}, and \protect\cite{FAE94}, respectively. 
The solid and dotted curves are from transport model calculations 
with free and dropping antiproton mass, respectively.
The experimental data from ref. \protect\cite{KIEN} for Ni+Ni collisions
and from ref. \protect\cite{JINR} for C+Cu collisions are shown 
by solid circles.}
\label{fig_antiproton}
\end{figure} 

The importance of medium effects on the nucleon mass can be seen from 
the significant decrease of the Q-value for the reaction $NN\to NNN\bar N$ 
in nuclear medium as a result of the attractive scalar potential.
This effect has been included in a number of studies based on transport
models.  In Fig. \ref{fig_antiproton}, theoretical results from these 
calculations for the antiproton differential cross section in Ni+Ni 
collisions at 1.85 GeV/nucleon \cite{antip,LI94C,MOS94} and C+Cu collisions 
at 3.65 GeV/nucleon \cite{FAE94} are compared with the experimental data 
from GSI \cite{KIEN} and Dubna \cite{JINR}. In \cite{antip,LI94C,MOS94}, 
the relativistic transport model has been used with the antiproton 
mean-field potential obtained from the Walecka model. The latter has a 
value in the range of -150 to -250 MeV at normal nuclear matter
density.  The results of ref. \cite{FAE94} are based on the nonrelativistic 
Quantum Molecular Dynamics with the produced nucleon and antinucleon masses 
taken from the Nambu$-$Jona-Lasinio model.  Within this framework, these 
studies thus show that 
in order to describe the antiproton data from heavy-ion collisions at 
subthreshold energies it is necessary to include the reduction of both 
nucleon and antinucleon masses in nuclear medium.  Even at AGS energies, 
which are above the antiproton production threshold in NN interaction, 
the medium effects on antiproton may still be important \cite{KKB91}. Indeed,
a recent study using the Relativistic Quantum Molecular Dynamics shows 
that medium modifications of the antiproton properties are important for a 
quantitative description of the experimental data \cite{RQMD95}.
We note, however, a better understanding of antiproton annihilation
is needed, as only about 5-10\% of produced antiprotons can survive,
in order to determine more precisely the antiproton in-medium mass
from heavy ion collisions.
 
\section{RESULTS FROM LATTICE QCD CALCULATIONS} 
\label{lattice}
Another source of information about possible in medium changes are lattice
QCD-calculations. For a review see e.g. \cite{DeT95}. Presently lattice
calculations can only explore systems at finite temperature but vanishing 
baryon chemical potential. Therefore, their results are not directly applicable
to present heavy ion experiments, where system at finite baryon chemical
potential are created. But future experiments at RHIC and LHC might succeed in
generating a baryon free region. But aside from the experimental aspects,
lattice results provide an important additional source where our model
understanding can be tested. Lattice calculation are usually carried out in
Euclidean space, i.e. in a space with imaginary time. As a consequence plane
waves in Minkowski-space translate into decaying exponentials in Euclidean
space. At zero temperature, masses of the low lying hadrons are extracted 
from two point functions, which carry the
quantum numbers of the hadron under consideration. 
\be
C(\tau) = \int d^3 x < J(x,\tau)\, J(0) > \sim \sum_i \alpha_i^2.
\exp(- m_i \tau)
\ee
Here, the sum goes over all hadronic states which carry the quantum numbers of
the operator $J$. At large imaginary times $\tau$ only the state with the 
lowest mass survives, and its mass can be determined from the exponential 
slope.
Zero temperature lattice calculation by now reproduce the hadronic spectrum to
a remarkable accuracy \cite{BCS94}. If one wants to extract masses at finite 
temperature, however, things become more complicated. The reason is that 
finite temperature on the lattice is equivalent to requiring periodic or 
anti-periodic boundary conditions in the imaginary time direction for bosons 
or fermions, respectively. Therefore, one cannot study the correlation 
functions at arbitrary large $\tau$ but is restricted to $\tau_{max} = 1/T$, 
where $T$ is the temperature under consideration. As a result, the lowest 
lying states cannot be projected out so easily. This has led people to 
study the correlation functions as a function of the spatial distance, 
where no restriction to the spatial extent exist, to extract so-called 
screening masses. At zero temperature the Euclidean time and space 
direction are equivalent and  the screening masses are identical to the 
actual hadron masses. By analyzing the correlation function in the spatial 
direction one essentially measures the range of a virtual particle emission. 
As already demonstrated by Yukawa, who deduced the mass of the pion from 
the range of the nuclear interaction, this can be used to extract particle 
masses. Screening masses, however, are not very useful if one wants to 
study hadronic masses at high temperature, close to $T_c$. At these 
temperatures, the spatial correlators are dominated by the trivial 
contribution from free independent quarks, and the resulting screening
masses simply turn out to be \cite{EI88}
\be
M_{\rm screen} = n \pi T,
\ee
where $n$ is the number of quarks of a given hadron, i.e. $n=2$ for mesons and
$n=3$ for baryons. Only for the pion and sigma meson, lattice calculation
found a significant deviation from this simple behavior, which is due to
strong residual interactions in these channel. All other hadrons considered so
far, such as the $\rho$ and $a_1$ as well as the nucleon, exhibit the above
screening mass \cite{DK87b,BGI91}. These trivial contributions from 
independent quarks are absent, however, in the time-like correlator. There 
has been one attempt to extract meson masses from the time-like correlators 
in four flavor QCD \cite{BGK95}. In this case it has been found that the mass 
of the $\rho$-meson does not change significantly below $T_c$. Above the 
critical temperature, on the other hand, the correlation function appeared 
to be consistent with one of noninteracting quarks, indicating the onset 
of deconfinement.

There have been attempts to extract the mass of the scalar $\sigma$
meson as well as that of the pion from so called susceptibilities \cite{Lea96}.
These are nothing else than the total four-volume integral over the two-point
functions. If the two-point function is dominated by one hadronic pole, 
then this integral should be inversely proportional to the square of the 
mass of the lightest hadron under consideration. Using this method, the 
degeneracy of the pion and sigma mass close to $T_c$ has been demonstrated.   

Finally, let us note that a scenario where all hadron masses scale linearly
with the chiral condensate is consistent with strong rise in the energy and
entropy  density around the phase transition as 
observed in Lattice calculations \cite{KB93b}. 

\section{SUMMARY}

In this review, hadron properties, particularly their masses, in the 
nuclear medium have been discussed.  Both conventional many-body 
interactions and genuine vacuum effects due to chiral symmetry
restoration have effects on the hadron in-medium properties. 

For Goldstone bosons, which are directly linked to the spontaneously
broken chiral symmetry, we have discussed the pion, kaons, and 
etas. For the pion, its mass is only slightly shifted in the medium
due to the small s-wave interactions as a result of chiral symmetry. 
On the other hand, the strong attractive p-wave interactions due
to the $\Delta(1230)$ and the $\rho(770)$  lead to a softening of 
the pion dispersion relation in medium. These effects result from
many-body interactions rather from chiral symmetry. Phenomenologically, 
the small change in the pion in-medium mass does not seem to have any
observable effects. However, the effects due to the softened pion in-medium 
dispersion relation may be detectable through the enhanced low transverse 
momentum pions in heavy ion collisions. 

For kaons, medium effects due to the Weinberg-Tomozawa
vector type interaction, which are negligible for pions
in asymmetric nuclear matter, are important
as they have only one light quark. However, because of the large 
explicit symmetry breaking due to the finite strange quark mass higher order
corrections in the chiral expansion are non-negligible,
leading to an appreciable attractive scalar interaction for both
the kaon and the antikaon in the medium. Available experimental 
data on both subthreshold kaon production and kaon flow are
consistent with the presence of this attractive scalar interaction.

For etas, their properties are more related to the explicit breaking 
of the $U_A(1)$ axial symmetry in QCD. If the $U_A(1)$ symmetry is restored 
in medium as indicated by the instanton liquid model, then their masses
are expected to decrease as well. 
Heavy ion experiments at the CERN-SPS, however, rule out a vast enhancement of
the final-state eta yield.  

In the case of vector mesons, we have discussed the rho, omega, and phi.
At finite temperature, all model calculations agree with the current
algebra result that the mass of the rho does not change to order $T^2$.
To higher order in the temperature and at finite density 
QCD sum rules predict a dropping of the masses in the medium. 
Predictions from chiral models, on the other hand, tend 
to predict an increase of the rho mass with temperature. 
However, presently available dilepton data from CERN SPS
heavy ion experiments are best described assuming a dropping rho mass
in dense matter but the errors in the present experimental data are too 
large to definitely exclude some more conventional explanations. 

For baryons, particularly the nucleon, the change of their properties
in the nuclear medium has been well-known in studies based on 
conventional many-body theory. On the other hand, the nucleon mass
is also found to be reduced in the medium as a result of the scalar 
attractive potential related to the quark condensate. A clear separation
of the vacuum effects due to the condensate from those of many-body
interactions is a topic of current interest and has not yet been resolved.
A dropping nucleon effective mass in medium seems to be required to
explain the large enhancement of antiproton production in heavy ion 
collisions at subthreshold energies.

In principle lattice QCD calculations could help answer quite a few of 
these questions, although they are restricted to systems at finite temperature
and vanishing baryon density. However, with the presently available computing
power reliable quantitative predictions about in-medium properties of hadrons 
are still not available. 

The study of in-medium properties of hadrons is a very active field, both
theoretically and experimentally. While many questions are still open and 
require additional measurements and more careful calculations, it is undisputed
that the study of in medium properties of hadrons provides us with an unique
opportunity to further our understanding about the long range, nonperturbative
aspects of QCD. 

\bigskip

{\bf Acknowledgments:} We are grateful to many colleagues for helpful 
discussions over the years. 
Also, we would like to thank G. Boyd, G. E. Brown, A. Drees, B. 
Friman, C. Gaarde, F. Klingl, R. Rapp, C. Song, T. Ullrich, and W. Weise
for the useful information and discussions during the preparation of this 
review. The work of CMK 
was supported in part by the National Science Foundation under Grant No. 
PHY-9509266. V.K. was supported the Director, 
Office of Energy Research, Office of High Energy and Nuclear Physics, 
Division of Nuclear Physics, and by the Office of Basic Energy
Sciences, Division of Nuclear Sciences, of the U.S. Department of Energy 
under Contract No. DE-AC03-76SF00098.
GQL was supported by the Department of Energy under Contract 
No. DE-FG02-88Er40388.

\end{document}